\documentclass[twocolumn,floatfix,superscriptaddress,a4paper,showkeys,nofootinbib,reprint]{revtex4-1}
\usepackage{epsfig}
\usepackage{latexsym}
\usepackage{xspace}
\usepackage[colorlinks=true,linktocpage=true,linkcolor=blue,citecolor=blue,allcolors=blue]{hyperref}
\usepackage[utf8]{inputenc}
\usepackage{indentfirst}
\usepackage{enumerate}
\usepackage{color}

\usepackage[caption=false,position=top]{subfig}

\usepackage{amsmath}
\usepackage{amssymb}
\usepackage[english]{babel}
\usepackage{url}

\usepackage{array}  
\usepackage{multirow}

\newcommand{\eq}[1]{\begin{align} #1 \end{align}}

\newcommand{\eqb}[0]{$b_{pn}=b_n$}
\newcommand{\neqb}[0]{$b_{pn}\neq b_n$}

\begin{document}

\title{
Quantum van der Waals quarkyonic matter at non-zero isospin asymmetry
}

\author{Tripp~Moss}
\affiliation{Physics Department, University of Houston, Box 351550, Houston, TX 77204, USA}

\author{Roman~Poberezhniuk}
\affiliation{Physics Department, University of Houston, Box 351550, Houston, TX 77204, USA}
\affiliation{Bogolyubov Institute for Theoretical Physics, 03680 Kyiv, Ukraine}
\affiliation{Frankfurt Institute for Advanced Studies, Giersch Science Center, Ruth-Moufang-Str. 1, D-60438 Frankfurt am Main, Germany}

\author{Volodymyr~Vovchenko}
\affiliation{Physics Department, University of Houston, Box 351550, Houston, TX 77204, USA}

\begin{abstract}

We extend the recently developed quantum van der Waals quarkyonic matter to non-zero isospin asymmetries by utilizing the two-component van der Waals equation with a generalized excluded volume prescription. The isospin dependence of van der Waals interaction parameters is determined by constraints on the symmetry energy, slope of the symmetry energy, and nuclear ground state properties. 
We find that the speed of sound has a peak for all values of the asymmetry parameter, signifying a transition to quarkyonic matter. The quarkyonic matter onset density is found to have a mild dependence on isospin asymmetry, 
with specific details influenced by the isospin dependence of the repulsive interactions.
We also incorporate leptonic degrees of freedom and explore the neutron star matter equation of state, calculating mass-radius relations and tidal properties of neutron stars. We find that quarkyonic matter supports heavy neutron stars with a maximum mass of at least 2.6 solar masses. 
We observe quantitatively different behavior for the excluded volume cases of isospin-blind ($b_{n}=b_{pn}$) and isospin-dependent ($b_{n} \neq b_{pn}$) 
repulsion, the latter being preferred by observational constraints.

\end{abstract}

\keywords{quarkyonic matter, quark-hadron duality, neutron star matter}

\maketitle

\section{Introduction}
\label{sec:intro}

Neutron star measurements provide valuable insights into the cold and dense regime of QCD that is inaccessible to first-principles calculations.
One can characterize the equation of state of dense QCD matter by the density dependence of the squared sound velocity, $v^2_s$. 
At extremely high densities, $v^2_s$ is expected to approach the conformal limit, $v^2_s\rightarrow 1/3$, as predicted by perturbative QCD~\cite{Kurkela:2009gj}. 
At the same time, neutron star observations suggest that the speed of sound may exhibit a pronounced peak significantly exceeding the conformal limit at densities several times nuclear saturation density \cite{Annala:2019puf,Altiparmak:2022bke,Brandes:2022nxa}, though this conclusion may not be definitive yet~\cite{Mroczek:2023zxo}. 
One appealing possibility here is that the peak reflects a transition to quarkyonic matter \cite{McLerran:2018hbz}, 
corresponding to quark degrees of freedom filling up the Fermi sea while confined baryonic excitations persist at the Fermi surface~\cite{McLerran:2007qj}.

The quarkyonic transition can be modeled phenomenologically by a mixture of effective baryonic and quark degrees of freedom with the fraction of quarks increasing with density. The effects of the Pauli exclusion principle are taken into account, restricting the baryon and quark momentum distributions. 
Namely, at zero temperature, a shell structure in momentum space emerges where deconfined quarks occupy the Fermi sea, and baryonic excitations appear only near the Fermi surface. 
Earlier versions of this model apply a phenomenologically motivated ansatz for the density evolution of the baryon Fermi surface thickness, as introduced in~\cite{McLerran:2018hbz,Zhao:2020dvu,Margueron:2021dtx}.

In Ref.~\cite{Jeong:2019lhv}, 
a dynamical mechanism for generating the momentum shell structure was proposed, 
namely, the quark fraction at a given baryon density is determined through energy minimization.
Introducing repulsive excluded volume interactions makes it energetically favorable to fill the low-momentum part of the phase space by quarks rather than nucleons, giving rise to the quarkyonic shell structure.
This mechanism has been used in a number of applications of quarkyonic matter in recent years~\cite{Duarte:2020xsp,Sen:2020qcd,Duarte:2020kvi,Duarte:2021tsx,Pang:2023dqj}, but also critically discussed in Ref.~\cite{Koch:2022act} where it was indicated that the inverse -- baryquark matter configuration -- may be energetically preferred in the quasi-particle picture.
Recent developments address this issue by utilizing explicit quark-hadron duality to obtain the quarkyonic
momentum shell structure~\cite{Kojo:2021ugu,Fujimoto:2023mzy,Fujimoto:2024doc}, although these descriptions currently indicate a singular behavior in the speed of sound at the onset of quarkyonic matter.
The onset of quarkyonic matter is typically regarded to take place at $n_B \gtrsim 2 n_0$, although recently it has been suggested that quarkyonic matter may occur close to nuclear saturation density of $n \simeq n_0 = 0.16$~fm$^{-3}$~\cite{Koch:2024qnz,McLerran:2024rvk}, indicating a possibility that ordinary nuclear matter may, in fact, be quarkyonic. 

In Ref.~\cite{Poberezhnyuk:2023rct}, a Quantum van der Waals~(QvdW) framework for quarkyonic matter was developed.
It is based on the quasi-particle picture and the excluded volume mechanism for generating the momentum shell structure but also incorporates an attractive mean-field, allowing one to describe the basic properties of symmetric nuclear matter and the associated nuclear liquid-gas transition.
In particular, the interaction parameters have been fixed to ground state properties of nuclear matter, allowing one to \emph{predict} the transition density to quarkyonic matter, which, depending on the excluded volume mechanism employed, occured at $n_B \simeq 1.5-2 n_0$.

Here we extend the QvdW quarkyonic matter framework to non-zero isospin flavor asymmetry. 
We calibrate the isospin dependence of the QvdW interaction parameters by using empirical constraints on the symmetry energy and its
slope.
We then analyze the dependence of equation of state and quarkyonic transition properties on the value of the charge fraction, $y = Q/B$, ranging from $y = 0.5$~(symmetric nuclear matter) to $y = 0$~(pure neutron matter). 
By incorporating leptonic degrees of freedom and the conditions of beta-equilibrium, we then apply the developed framework to explore the neutron-star matter equation of state.
The resulting mass-radius relations indicate that quarkyonic matter can support heavy neutron stars up to $M \simeq 2.6 M_{\odot}$, and aligns with available observational constraints when empirical estimates of the symmetry energy slope are reproduced.

The paper is organized as follows.
Section \ref{sec:Pauli} describes the generalized quarkyonic matter setup for an arbitrary value of isospin asymmetry.
In Sec.~\ref{sec:interaction}, we implement different scenarios for interactions between nucleons and constrain the interaction parameters from the known low-density properties of nuclear matter.
Section~\ref{sec:y-dependence} presents the dependence of quarkyonic matter properties as a function of isospin asymmetry.
In Sec.~\ref{sec:NS} we include leptons and apply the resulting equation of state for the description of neutron stars.
Summary and discussion in Sec.~\ref{sec:summary} close the article.

\section{Mixture of nucleons and quarks with Pauli exclusion principle}
\label{sec:Pauli}

We model nuclear matter as a mixture of quarks and nucleons at zero temperature.
To allow for non-zero isospin asymmetry, we consider protons~(p), neutrons~(n), $u$ and $d$ quarks as explicit degrees of freedom.
Consider first a mixture of non-interacting particles.
At zero temperature, $T=0$, baryons and quarks occupy all momentum states up to the Fermi levels $k_{p,u}$ and $q_{u,d}$, respectively. 
The densities of baryon number, electric charge, and energy contain contributions from quarks and baryons and read\footnote{Here $n_{u,d}^{\rm id}$ are summed over all $N_c$ colors, hence they carry integer contribution to baryon and electric charge number.}
\eq{
\rho_B & = n_p^{\rm id}(k_F^p) + n_n^{\rm id}(k_F^n) + n_u^{\rm id}(q_F^u) + n_d^{\rm id}(q_F^d), \\
\rho_Q & = n_p^{\rm id}(k_F^p) + 2 n_u^{\rm id}(q_F^u) - n_d^{\rm id}(q_F^d), \\
\label{eq:en}
\varepsilon & = \varepsilon_p^{\rm id}(k_F^p) + \varepsilon_n^{\rm id}(k_F^n) + \varepsilon_u^{\rm id}(q_F^u) + \varepsilon_d^{\rm id}(q_F^d).
}

With the ideal gas functions given as
\eq{\label{eq:nN}
n_{p,n}^{\rm id}(k_F^{p,n})&=\frac{g}{2\pi^2}\int_{0}^{k_F^{p,n}}k^2 \, \rho_{p,u}(k) dk, \\
\label{eq:eN}
\varepsilon_{p,n}^{\rm id}(k_F^{p,n})&=\frac{g}{2\pi^2}\int_{0}^{k_F^{p,n}}k^2 \, \rho_{p,n} (k) \, \epsilon_{p,n}(k) dk, \\\label{eq:nq}
n_{u,d}^{\rm id}(q_F^{u,d})&=\frac{g}{2\pi^2}\int_{0}^{q_F^{u,d}}q^2 \, \rho_{u,d}(q) dq, \\\label{eq:eq}
\varepsilon_{u,d}^{\rm id}(q_F^{u,d})&=N_c\frac{g}{2\pi^2}\int_{0}^{q_F^{u,d}} q^2 \, \rho_{u,d}(q) \, \epsilon_{u,d}(q) dq.
}
Here $\rho_{p,n}(k),\rho_{u,d}(q)$ are the modification factors for the phase space densities of baryon and quark states due to the Pauli exclusion principle (explained below),
$\epsilon_{u,d}(q)= \sqrt{m_{u,d}^2+q^2}$ and
$\epsilon_{p,n}(k)= \sqrt{m_{p,n}^2+k^2}$
are relativistic single-particle energies.
We take $m_p = m_n = N_c m_u = N_c m_d = 938$~MeV/$c^2$.
$g = g_{u,d} = g_{p,n} = 2$ is the spin degeneracy factor for quarks and nucleons.
Note that the quarks have an additional degeneracy factor of $N_c$ due to color.
This factor is canceled out by the $1/N_c$ fractional baryon charge in the expressions~\eqref{eq:nq} for quark contribution to the baryon and electric charge densities  but is present in Eq.~\eqref{eq:eq} describing quark energy density.

The Fermi levels $k_F^{p,n}$, $q_F^{u,d}$ as well as the functions $\rho_{p,n}(k), \rho_{u,d}(q)$ define the values of baryon and electric charge densities, $\rho_B$ and $\rho_Q$. Alternatively, the true equilibrium configuration of the system for given values of $\rho_B$ and $\rho_Q$ is determined by finding the values of Fermi momenta which uniquely minimize the energy density $\varepsilon$. 

The functions $\rho_{p,n}(k)$ and $\rho_{u,d}(q)$ modify the density of states for baryons and quarks such that the Pauli exclusion principle between confined and deconfined quarks is respected.
In the absence of the Pauli exclusion principle, one would have $\rho_{p,n}(k) = 1$ and $\rho_{u,d}(q) = 1$.
In a more realistic setup, one must account for Pauli blocking between confined and deconfined quarks, namely that they cannot both occupy the same momentum  state~\cite{McLerran:2018hbz,Jeong:2019lhv}.

 In quarkyonic matter the baryonic excitations form a momentum shell near the Fermi surface, while the quarks fill lower momentum states in the Fermi sea. It is motivated by the notion that small relative momentum interactions associated with confinement are blocked in the ``bulk" Fermi sea of quarks, up to the momentum $q_{\rm bu} = k_{\rm bu} / N_c$. Thus as density increases, confinement can only occur near the Fermi surface, leading to the dynamical generation of the shell of baryons with a thickness $\Delta = k_F - k_{\rm bu}$.  

Let us formulate the Pauli exclusion principle of quarkyonic matter for the general case of asymmetric nuclear matter.
First, we introduce the following asymmetry parameter
\eq{
y = \frac{\rho_Q}{\rho_B},
}
which has the meaning of the charge~(or proton) fraction.

The realization of the Pauli principle in quarkyonic matter has been studied extensively for the case of symmetric matter, $y = 1/2$.
In this case, the  Fermi surfaces of isospin partners ($u,d$ and $p,n$) are degenerate, resulting in momentum distribution functions which read
\eq{
\rho_{u}^{y = 1/2} (q) = \rho_{d}^{y = 1/2} (q) & = \frac{\sqrt{q^2 + \Lambda^2}}{q}\, \Theta(q_{\rm bu} - q), \\
\rho_{p}^{y = 1/2} (k) = \rho_{n}^{y = 1/2} (k) & = \Theta(k - k_{\rm bu}) \Theta(k_F - k),
}
with $k_{\rm bu} = N_c q_{\rm bu}$.
Here $\Lambda$ is the infrared regulator, introduced to smooth out the sudden onset of quarks characteristic of quarkyonic matter~\cite{Jeong:2019lhv}, by modifying the density of states at low momenta relative to that given by free constituent quark limit.
The value of $\Lambda$ regulates the sharpness of the peak in the density dependence of $v_s^2$ but has small influence on the baryon density where this peak appears.
The value of $\Lambda$ is not very well constrained, with recent Bayesian analysis in Ref.~\cite{Pang:2023dqj} indicating $\Lambda \sim 300 \pm 100$~MeV. 
In our previous work we used a value $\Lambda = 200$~MeV and we preserve this value here for consistency.
The parameters $k_{\rm bu}$ and $k_F$ are determined through energy minimization at fixed baryon density $\rho_B$.

To describe asymmetric matter, let us first consider the extreme case of neutral (pure neutron) matter, $y = 0$.
We shall require that electric charge neutrality applies separately in both the nucleon and quark sectors of the theory.
Although such a choice is, in principle, not unique, it is well-motivated by considering the duality between quarks and nucleons consisting of quarks.
In this case, one has $n_d = 2 n_u$ and $n_p = 0$.
The Fermi shell of neutrons starts from momentum $k_{\rm bu} = N_c q_{\rm bu}^d$ as the lower momenta are blocked due to saturation of $d$ quarks.

The Fermi surfaces of $u$ and $d$ quarks are determined by momenta $q_{\rm bu}^u$ and $q_{\rm bu}^d$, respectively.
It is an open question how the Fermi surface of $u$ quarks is related to that of $d$ quarks. 
From the point of view of energy minimization, one would fill the Fermi sea of $u$ quarks up to a lower Fermi momentum $q_{\rm bu}^u = 2^{-1/3} q_{\rm bu}^d$ to maintain the condition $n_d = 2 n_u$. 
This was the framework adopted e.g. in~\cite{McLerran:2018hbz}.
On the other hand, if one considers the quark-hadron duality, this would lead to unnatural momentum distribution asymmetry between $u$ and $d$ quarks.
Therefore, here we adopt a different approach, first obtained in Ref.~\cite{Koch:2024qnz} within a dual quarkyonic model,
where the Fermi surfaces of $u$ and $d$ quarks coincide.
In this case $q_{\rm bu}^u = q_{\rm bu}^d = q_{\rm bu}$, but, due to charge neutrality, 
the $u$ quark levels are half-filled.
One obtains
\eq{
\rho_{u}^{y = 0} (q)& = \frac{1}{2} \frac{\sqrt{q^2 + \Lambda^2}}{q}\, \Theta(q_{\rm bu} - q), \\
\rho_{d}^{y = 0} (q)& = \frac{\sqrt{q^2 + \Lambda^2}}{q}\, \Theta(q_{\rm bu} - q),\\
\rho_{p}^{y = 0} (k) & = 0,\\
\rho_{n}^{y = 0} (k) & = \Theta(k - k_{\rm bu}) \Theta(k_F - k).
}
As before, $k_{\rm bu} = N_c q_{\rm bu}$, and $k_{\rm bu}$ and $k_F$ are determined through energy minimization at fixed baryon density $\rho_B$.

We can now generalize these considerations for arbitrary value of $y$ in the range $0 < y < 1/2$.
Requiring the charge fraction in the quark sector to be equal to $y$ leads to the following condition:
\eq{
\frac{2n_u - n_d}{ n_u + n_d} = y,
}
written here for $N_c = 3$. 
This entails
\eq{
n_d = \frac{2-y}{1+y} n_u.
}
Therefore, in the quark sector we have
\eq{
\rho_{u} (q)& = \frac{1+y}{2-y} \frac{\sqrt{q^2 + \Lambda^2}}{q}\, \Theta(k_{\rm bu}/N_c - q), \\
\rho_{d} (q)& = \frac{\sqrt{q^2 + \Lambda^2}}{q}\, \Theta(k_{\rm bu}/N_c - q),
}
for $0 \leq y \leq 1/2$, so that the quark contributions to the baryon density read
\eq{\label{eq:qy:nu}
n_u & = \frac{1+y}{2-y} \frac{g}{2\pi^2}  \int_0^{k_{\rm bu}/N_c} q \sqrt{q^2 + \Lambda^2} dq,\\
\label{eq:qy:nd}
n_d & = \frac{g}{2\pi^2} \int_0^{k_{\rm bu}/N_c} q \sqrt{q^2 + \Lambda^2} dq.
}

In the baryon sector we have
\eq{
\rho_{p,n} (k) & = \Theta(k - k_{\rm bu}) \Theta(k_F^{p,n} - k),
}
where $k_F^p$ and $k_F^n$ are constrained from the condition $n_p / (n_p + n_n) = y$, implying
\eq{\label{eq:qy:npn}
n_{p,n}^{\rm id} & = \frac{g}{2\pi^2}  \int_{k_{\rm bu}}^{k_F^{p,n}} k^2 dk = \frac{g}{6\pi^2} [(k_F^{p,n})^3 - k_{\rm bu}^3].
}

In practice, we perform calculations at fixed values of baryon density $\rho_B$, charge fraction $y$, and quark fraction $f_q = (n_u + n_d) / \rho_B$.
The algorithm is the following:
\begin{enumerate}
    \item Using Eqs.~\eqref{eq:qy:nu} and \eqref{eq:qy:nd} calculate the value of $k_{\rm bu}$ such that $n_u + n_d = f_q \rho_B$.
    \item Using Eq.~\eqref{eq:qy:npn} compute the values of $k_F^p$ and $k_F^n$ such that $n_p + n_n = (1-f_q) \rho_B$ and $n_p/(n_p+n_n) = y$. Note that the calculation of $n_p$ and $n_n$ should include the treatment of nuclear interactions described in the next section.
    \item Calculate the energy density $\varepsilon$ using Eq.~\eqref{eq:en}.
    \item To find the equilibrium configuration at given $\rho_B$ and $y$, vary $f_q$ in the range $[0,1]$ to find the minimum in energy density $\varepsilon$.
\end{enumerate}

The isospin asymmetric quark-baryon mixture can also be constructed for other momentum space configurations, such as baryquark~\cite{Koch:2022act} or uniform quarkyonic~\cite{Poberezhnyuk:2023rct} mixtures. These investigations are left for future work.

\section{Nuclear interactions}
\label{sec:interaction}

The introduction of interactions is necessary to dynamically generate the appearance of quarks at sufficiently high baryon densities.
This is so because the energy density of a free nucleon gas is always smaller than that of a free quark gas for the same value of baryon number density, as originally pointed out in~\cite{Jeong:2019lhv}.
Here we incorporate nuclear interactions while treating quark quasiparticles in the framework of an ideal gas with an infrared regulator,  $n_Q(q_F)=n_Q^{\rm id}(q_F)$. 
We also neglect quark-nucleon interactions apart from the Pauli exclusion principle described in the previous section.
We use a quantum van der Waals framework from Refs.~\cite{Vovchenko:2015vxa} with a generalized excluded volume prescription~\cite{Vovchenko:2017cbu}, 
which has earlier been applied to isospin symmetric quarkyonic matter~\cite{Poberezhnyuk:2023rct}.

\subsection{Quantum van der Waals model}

Let us start with the standard quantum van der Waals model.
The two-component version of the QvdW model for asymmetric nuclear matter has been developed in Refs.~\cite{Vovchenko:2017zpj,Poberezhnyuk:2018mwt}.
The proton and neutron number densities read
\eq{\label{eq:vdw-np}
n_p & = (1-b_n n_p - b_{pn} n_n) n_p^{\rm id}, \\\label{eq:vdw-nn}
n_n & = (1-b_{pn} n_p - b_{n} n_n) n_n^{\rm id}.
}
These can be written in compact form
\eq{
n_{p,n} = f_{\rm vdW} (x_{p,n}) n_{p,n}^{\rm id},
}
where
\eq{\label{eq:fev}
f_{\rm vdW} (x) & = 1 - x, \\
x_p & = b_n n_p + b_{pn} n_n, \\
x_n & = b_{pn} n_p + b_{n} n_n,
}
and the expression~\eqref{eq:qy:npn} for $n_{p,n}^{\rm id}$ is described in the previous section.
The energy density reads
\eq{\label{eq:envdw}
\varepsilon_N = \sum_{i \in p,n} f_{\rm vdW}(x_i) \varepsilon^{\rm id}_i - a_n (n_p^2 + n_n^2) - 2 a_{pn} n_p n_n.
}

Here $b_n$ and $a_n$ correspond to repulsive and attractive interactions among neutrons~(and, by isospin symmetry, among protons). On the other hand, $b_{pn}$ and $a_{pn}$ correspond to interactions in the proton-neutron isospin channel.

In the limit $y = 0$, the model reduces to a single-component QvdW equation for neutrons with vdW parameters $b_n$ and $a_n$.
On the other hand, in the symmetric matter limit, $y = 1/2$, the model reduces to a single-component QvdW equation for nucleons with spin-isospin degeneracy factor $4$ and symmetric matter QvdW parameters $a = (a_n + a_{pn})/2$ and $b = (b_n + b_{pn})/2$.

\subsection{Other real gas models}

The quantum van der Waals model can be expanded into a more general class of real gas models~\cite{Vovchenko:2017cbu}.
This can be achieved both through the modification of the excluded volume function $f(x)$ in Eq.~\eqref{eq:fev} and the mean field contributions to the energy density in Eq.~\eqref{eq:envdw}.
Here we only consider the former. Following Ref.~\cite{Poberezhnyuk:2023rct}, we explore Carnahan-Starling~\cite{Carnahan:1969} and trivirial~\cite{Vovchenko:2019hbc} models, which correspond to
\eq{
f_{\rm CS} (x) & = \exp\left[-\frac{3 x}{4 - x}-\frac{4 x}{(4 - x)^2}\right]~, \\
f_{\rm TVM} (x) & = \exp\left[-x-\frac{x^2}{2}\right].
}

All available models are known to overestimate considerably the empirical estimates for nuclear incompressibility factor $K_0$.
The density-dependent excluded volume prescriptions studied here improve the description of $K_0$ somewhat, but not sufficiently enough to reproduce the empirical constraints. 
This issue can be addressed by considering density-dependent mean fields~\cite{Vovchenko:2017cbu,Dutra:2020qsn}.
We leave these improvements of nuclear matter description for future work and instead focus on the transition to quarkyonic matter.
We note that, for this reason, it may be very prudent to use the slope of symmetry energy to constrain the interaction parameters, given that we have already overestimated compressibility. 
We, therefore, explore two scenarios for treating the symmetry slope constraints, as detailed in the following subsection.

\subsection{Constraining the interaction parameters}

\begin{figure}
    \includegraphics[width=.49\textwidth]{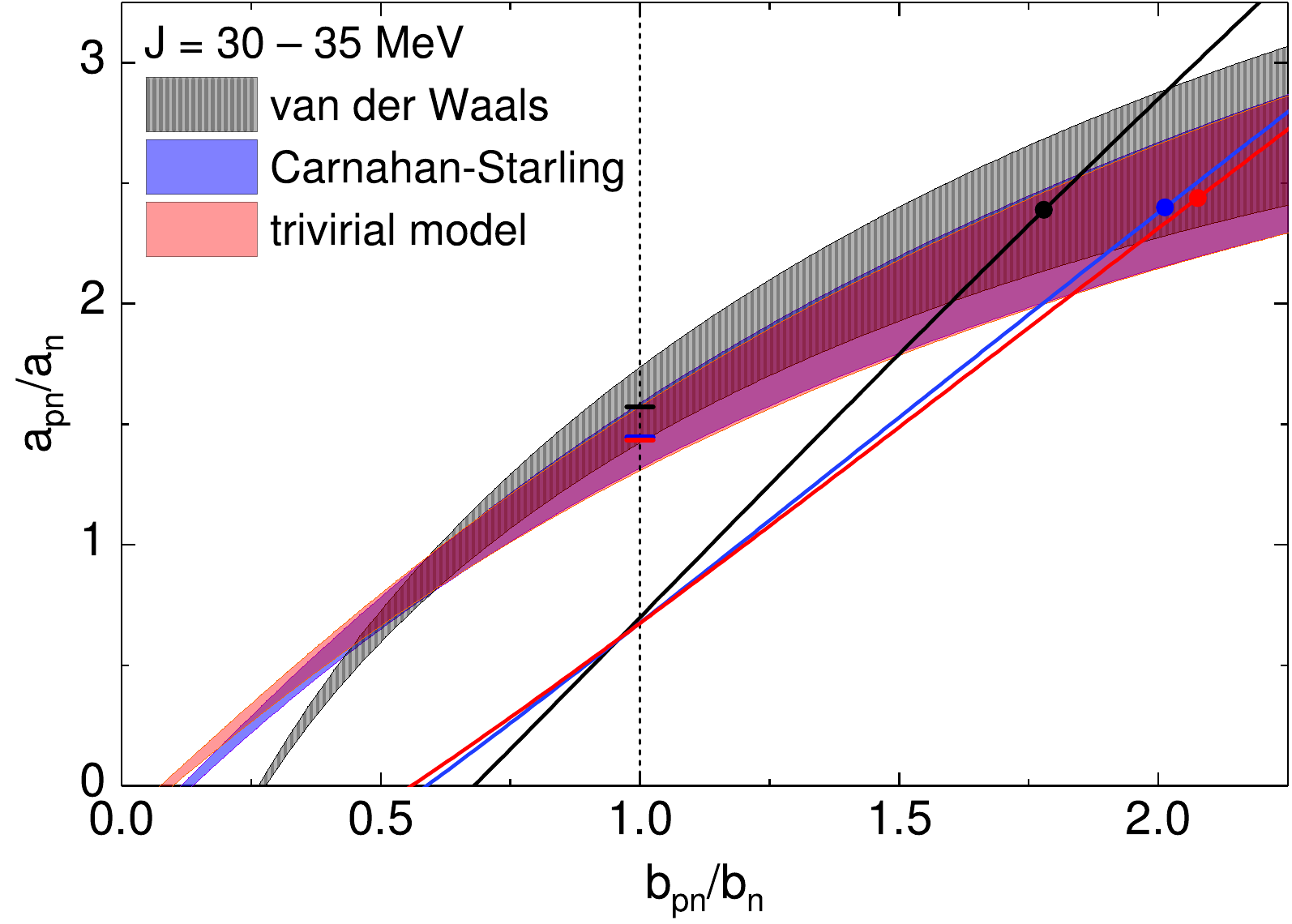}
    \caption{Contours of symmetry energy $J = 32.5 \pm 2.5$~MeV and lines of constant symmetry energy slope $L=58.9$ MeV in the plane of $b_{pn} / b_n$ and $a_{pn} / a_n$ for the van der Waals (grey) Carnahan-Starling (blue) and trivirial model (red) equation of state.  Horizontal dashes and full circles correspond to the chosen values of the parameters for, respectively, the case of $b_{pn}=b_{n}$ and $b_{pn}\neq b_{n}$.}
    \label{fig:abpn}
\end{figure}

The parameters $a = (a_{pn} + a_n) / 2$ and $b = (b_{pn} + b_n)/2$ have been fixed to reproduce the ground state of symmetric nuclear matter at $n_N = \rho_0 = 0.16$~fm$^{-3}$~\cite{Vovchenko:2015vxa,Vovchenko:2017cbu}. 
For instance, for vdW model this gives $a = 329$~MeV fm$^3$ and $b = 3.42$~fm$^3$~\cite{Vovchenko:2015vxa}.
To apply the model to asymmetric matter, we need to fix two additional parameters that probe the isospin dependence of interaction, for instance, ratios $a_{pn} / a_n$ and $b_{pn} / b_n$.
The simplest choice would be to set these ratios to unity, $a_{pn} / a_n = b_{pn} / b_n = 1$, but in this case, the model yields symmetry energy of $J \approx 20$~MeV, considerably below the empirical value of $J = 30-35$~MeV~\cite{Baldo:2016jhp}.
Figure~\ref{fig:abpn} shows the contour $J = 32.5 \pm 2.5$~MeV in the plane of $a_{pn} / a_n$ and $b_{pn} / b_n$.

\begin{center}
\begin{table}
\begin{tabular}
{ | l | c | c | c | c | c | c | }
\hline
 & \multicolumn{2}{c|}{~vdW~}  & \multicolumn{2}{c|}{~CS~}  & \multicolumn{2}{c|}{~TVM~}   \\
\hline
~$a~[\rm{MeV} ~ fm^3]$   & \multicolumn{2}{c|}{~329~} & \multicolumn{2}{c|}{~347~} & \multicolumn{2}{c|}{~349~} \\
~$b~[\rm fm^3]$          & \multicolumn{2}{c|}{~3.42~} & \multicolumn{2}{c|}{~4.43~} & \multicolumn{2}{c|}{~ 4.28~} \\
\hline
~$a_{pn}/a_n$            & 1.57 & 2.48 & 1.45 & 2.40 & 1.43 & 2.44\\
~$b_{pn}/b_n$            & 1   & 1.98  & 1    & 2.01 &  1   & 2.08\\
~$L$~[MeV]            & 122   & 58.9  & 121    &  58.9 &  121   &  58.9\\
\hline
\end{tabular}
\caption{
\label{tab:NM}
Values of the interaction parameters and the resulting symmetry energy slope, $L$ for three equations of state and two cases, $b_{pn}=b_{n}=b$ and $b_{pn}\neq b_{n}$. The value of $a_{pn}/a_n$ is fitted to reproduce the symmetry energy of $J=32.5$~MeV.
}
\end{table}
\end{center}

Along with symmetry energy, one needs an additional constraint to fix both $a_{pn} / a_n$ and $b_{pn} / b_n$ simultaneously.
Here we consider two possibilities: 
\begin{enumerate}
    \item We implement constraints on both the symmetry energy $S = 30-35$~MeV and its slope $L$~\cite{Baldo:2016jhp}. 
    We use $S = 32.5$~MeV for the symmetry energy constraint and take the value $L = 58.9$~MeV from~\cite{Li:2013ola} for the slope.
    We label this scenario as $b_{pn} \neq b_n$ and the resulting values of the parameters are listed in Table~\ref{tab:NM}.
    
    \item We omit the constraint on the symmetry energy slope $L$ and assume that the excluded volume repulsion is isospin-blind, namely $b_{pn} = b_n = b$. 
    This scenario can be motivated by the finding within the IdylliQ quarkyonic matter that the onset of quarkyonic matter in symmetric and pure neutron matter happens at essentially the same density~\cite{Koch:2024qnz}. 
    Of course, in such case, we may not have a good handle on the slope of symmetry energy, and the value of $L$ in this $b_{pn} = b_n$ scenario is $L \sim 121-122$~MeV, which is considerably above the constraint from Ref.~\cite{Li:2013ola}.
    However, empirical constraints on $L$ still vary significantly depending on the method used.
    The value $L \sim 121-122$~MeV is compatible with the estimate $L = 106 \pm 37$~MeV~\cite{Reed:2021nqk} based on PREX-II measurements~\cite{PREX:2021umo}.
    Furthermore, it may be more desirable to improve the description of nuclear compressibility in our approach first before incorporating the $L$ constraint.

\end{enumerate}

\section{Properties of Quarkyonic Matter as a function of isospin asymmetry}
\label{sec:y-dependence}

Now that we have established the framework for our calculations, we turn our attention to the quarkyonic model for pure nuclear matter. We investigate each excluded volume formalism, considering both the $b_{pn}=b_{n}$ and $b_{pn}\neq b_{n}$ scenarios. 
In all cases we calculate the values of the energy per baryon $\varepsilon/\rho_B - m_N$ with the subtraction of nucleon mass~(corresponding to the binding energy per baryon in the nuclear sector), quark fraction $f_q = (n_u + n_d) / \rho_B$, and the squared sound velocity $v_s^2$. Each of these quantities is investigated as a function of baryon density at constant charge fraction, and then the charge fraction is varied to analyze the dependence on $y$.

\begin{figure*}
    \includegraphics[width=.49\textwidth]{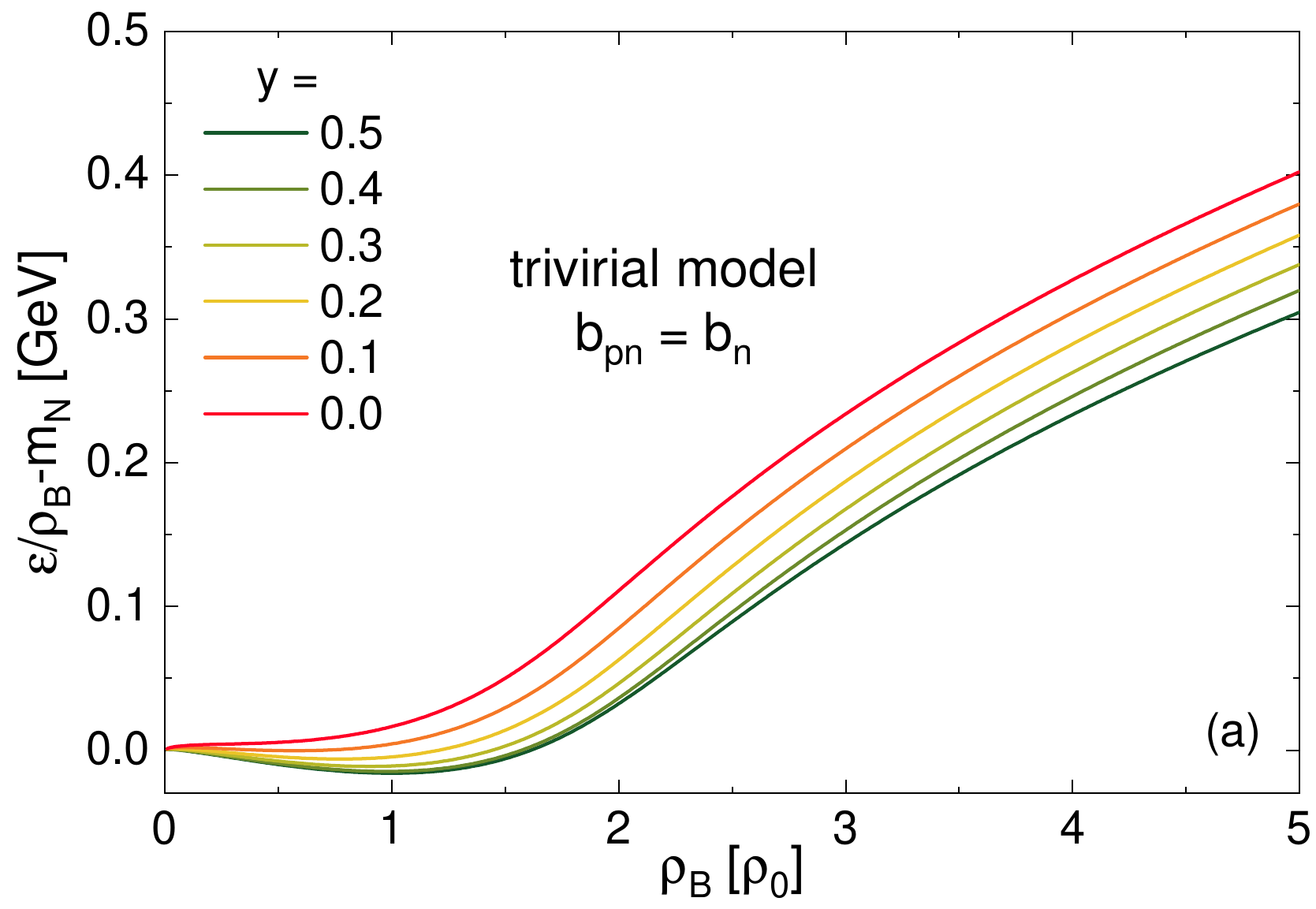}
    \includegraphics[width=.49\textwidth]{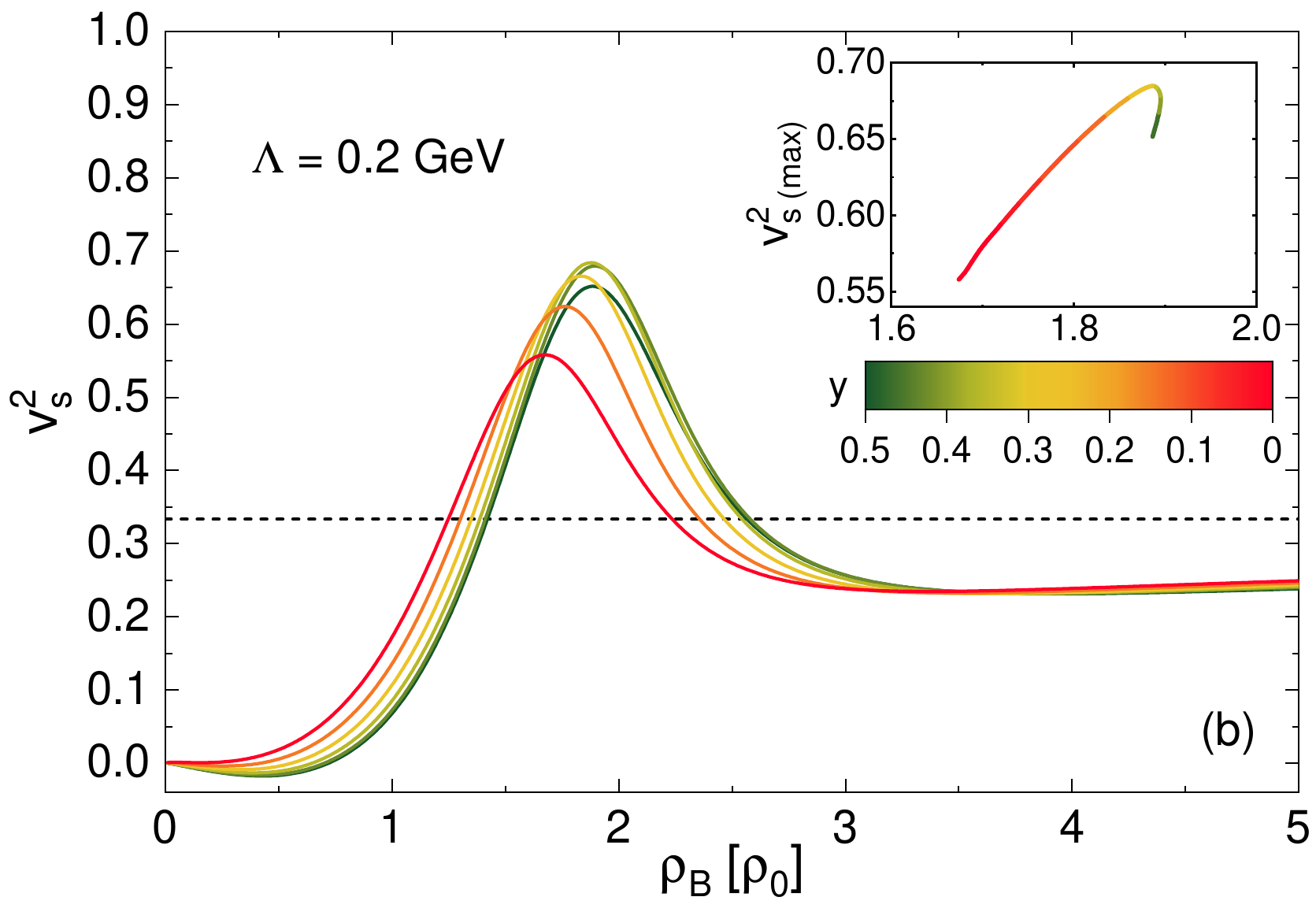}
    \includegraphics[width=.49\textwidth]{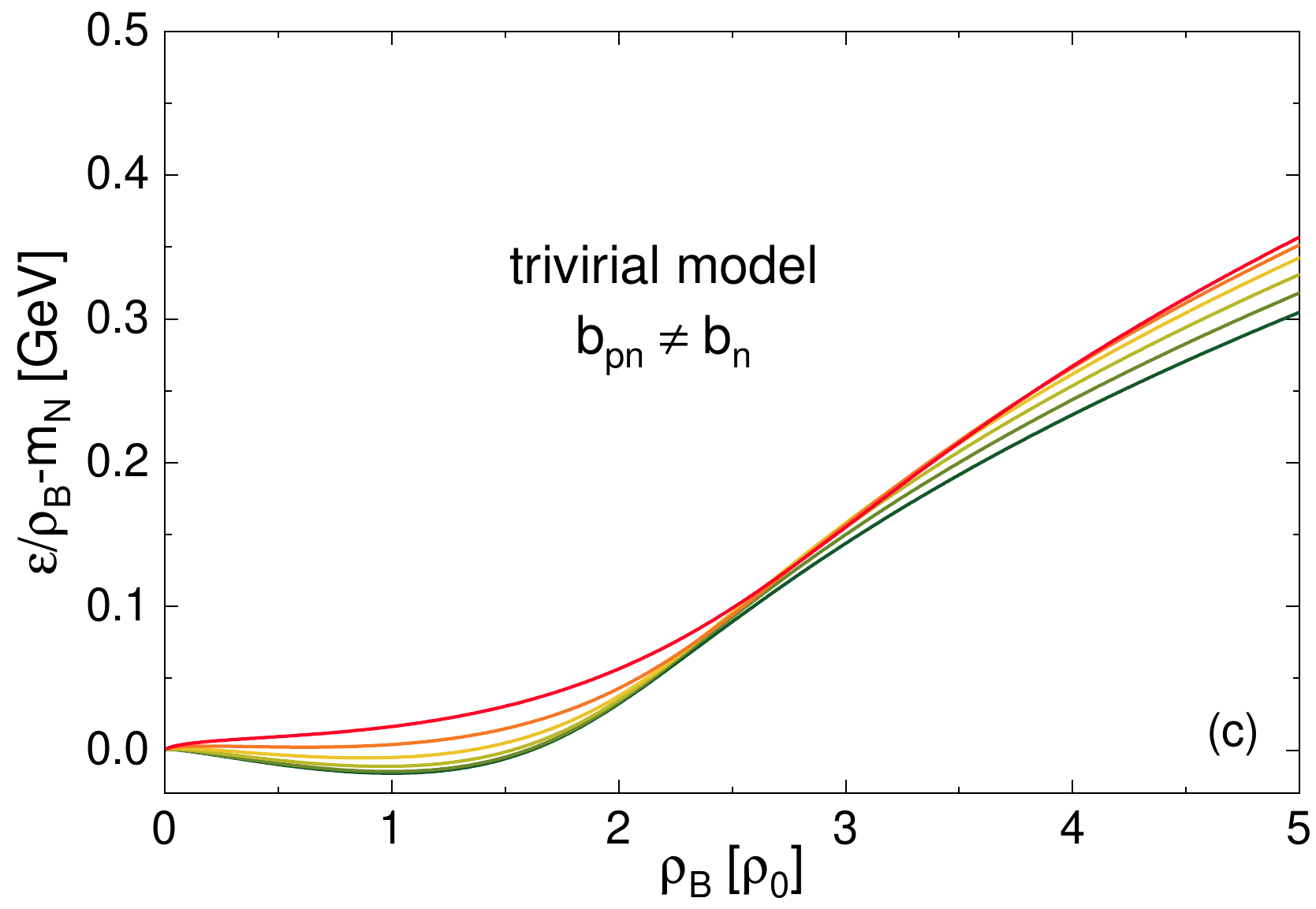}
    \includegraphics[width=.49\textwidth]{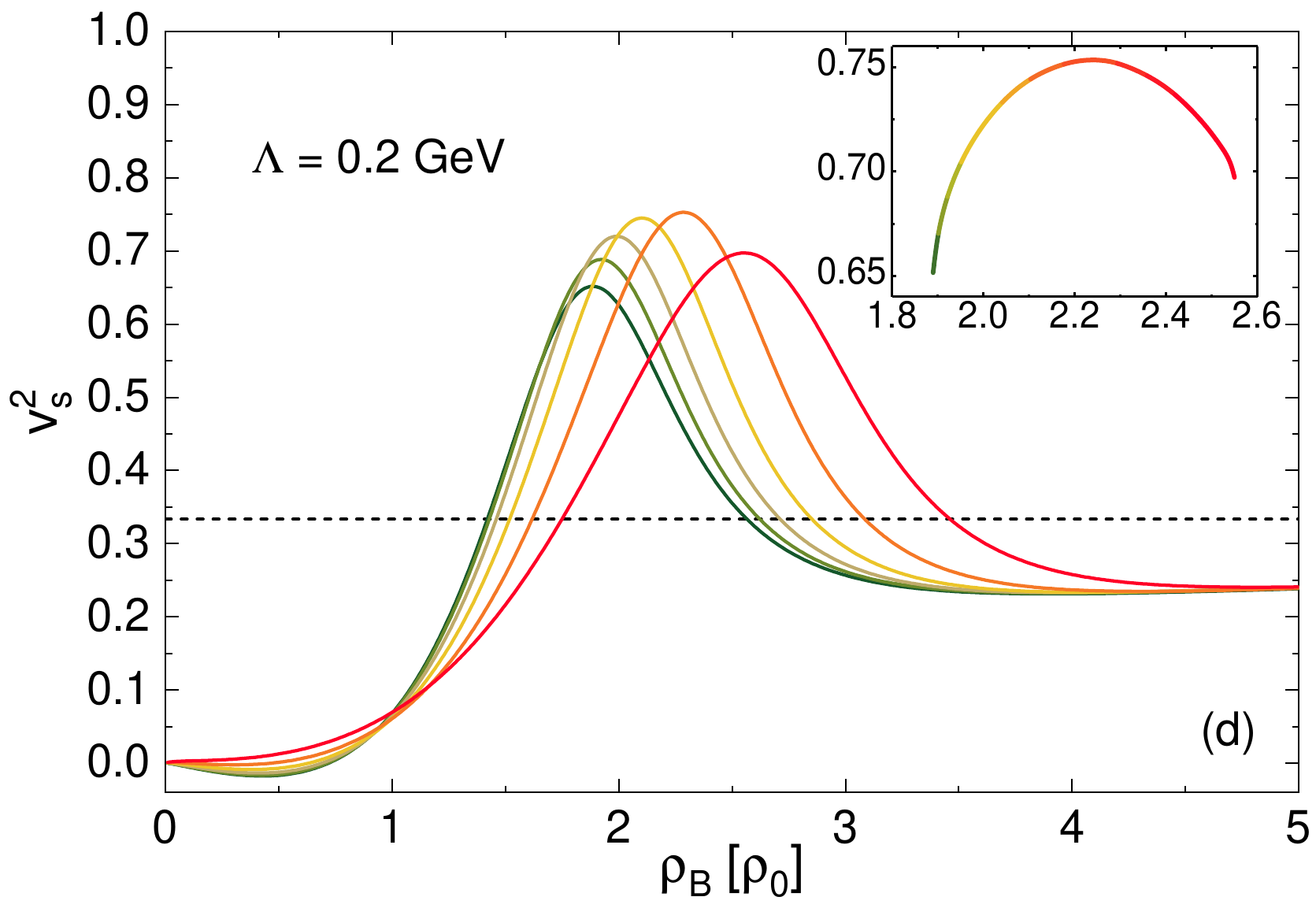}
    \caption{Binding energy per baryon (left) and sound velocity (right) as functions of baryon density for different values of the charge fraction $y$ from isospin-symmetric matter $y=0.5$ to pure neutron matter $y=0$.
    Trivirial model treatment of excluded volume is used as an example with isospin-blind repulsion, \eqb, in panels (a) and (b) and isospin-dependent repulsion, \neqb, in panels (c) and (d).
    The insets in panels (b) and (d) show the sound velocity at the peak $v^2_{s({\rm max})}$ versus the corresponding baryon density of the peak $\rho_B^{\rm tr}$ with charge fraction $y$ varying continuously between $y=0.5$ and $y=0$.} 
    \label{fig-y}
\end{figure*}

We note that the squared sound velocity $v_s^2$ offers a crucial tool for understanding the equation of state of dense matter; at zero temperature, it corresponds to
\eq{
v_s^2=\frac{\rho_B}{\mu_B}\frac{d \mu_B}{d \rho_B}=\frac{\rho_B}{\mu_B}\frac{d^2 \varepsilon}{d \rho_B^2}~.
}
A hallmark of the transition into the quarkyonic matter phase is often a pronounced peak in sound velocity, surpassing the conformal limit. This pattern appears consistently across all our computations, with the baryon density at the sound velocity peak acting as an indicator for the quarkyonic matter transition density $\rho_B^{tr}$.

We see qualitatively different results depending on our treatment of excluded volume. In the case of van der Waals when $b_{pn}\neq b_{n}$, $v_s^2$ shows a discontinuity between low proton content and pure neutron matter ($y=0$)
and violates the causality constraint.
This is because the quarkyonic transition occurs 
in the vicinity of the limiting close-packing density where $f_{\rm vdW} (x)=0$. Since $p-n$ repulsion is much stronger than $n-n$, adding 
even infinitesimal proton density
to pure neutron matter causes a large jump in the overall energy content of the system. This
issue does not arise when $b_{pn}=b_{n}$, as $p-n$ and $n-n$
repulsions in this case are equally strong.

For both Carnahan-Starling and trivirial model excluded volumes,
the speed of sound varies smoothly and remains casual across the entire range of isospin asymmetries.
Binding energy per baryon and speed of sound are shown in Fig.~\ref{fig-y} for the trivirial model as an example.
We do see different behavior in the peak of $v_s^2$ depending on the treatment of excluded volume parameters. When $b_{pn}\neq b_{n}$, the location of the $v_s^2$ peak
shifts monotonically to higher baryon densities as the charge fraction decreases.
For $b_{pn}=b_{n}$, the location of the peak shows non-monotonic behavior.
The energy density per baryon for different excluded volume treatments converges around the transition density when $b_{pn}\neq b_{n}$. In contrast, for $b_{pn}=b_{n}$ the energy density per baryon in neutron rich matter is noticeably higher than in symmetric matter at all considered baryon densities.

\begin{figure*}
    \includegraphics[width=.49\textwidth]{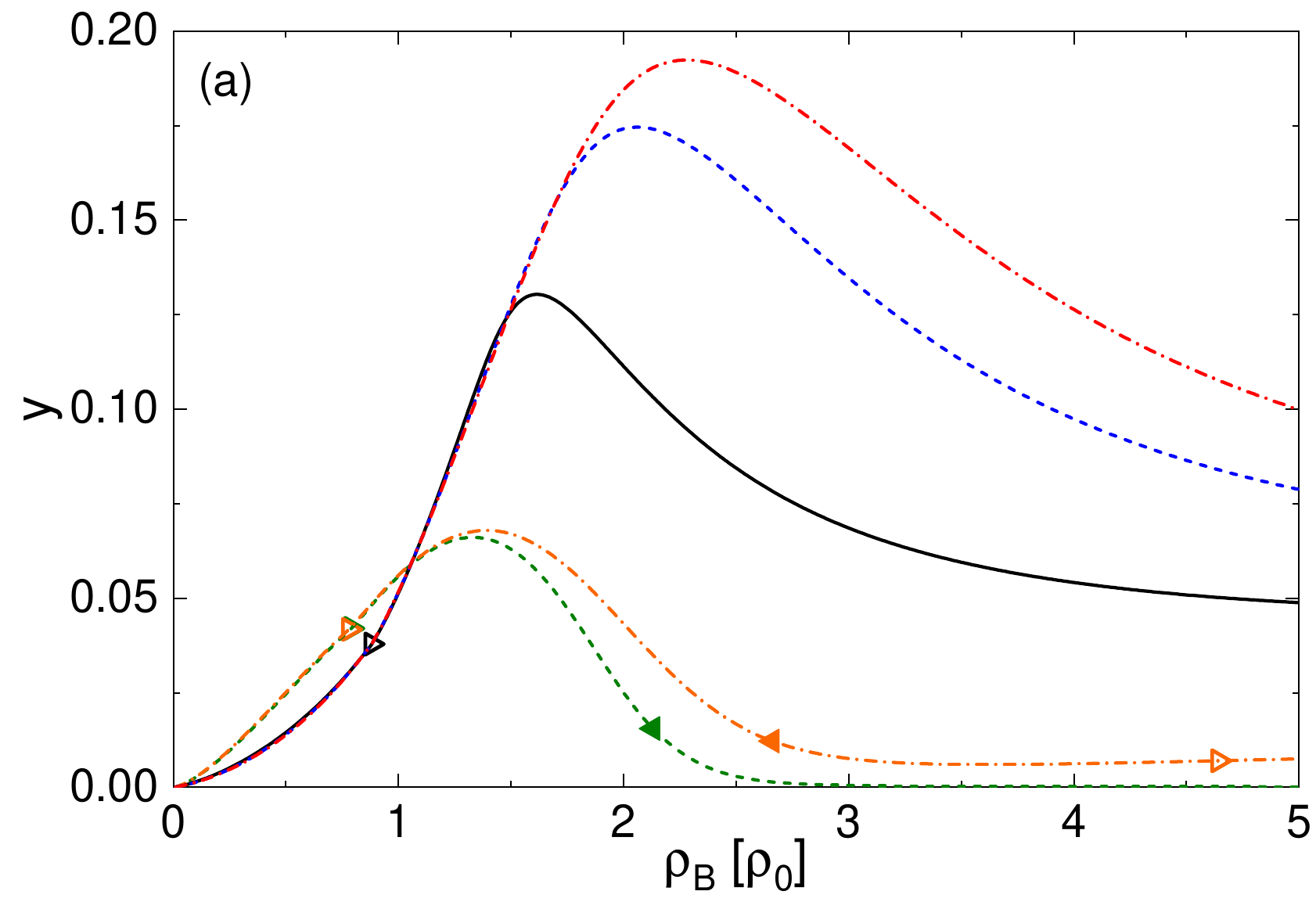}
    \includegraphics[width=.49\textwidth]{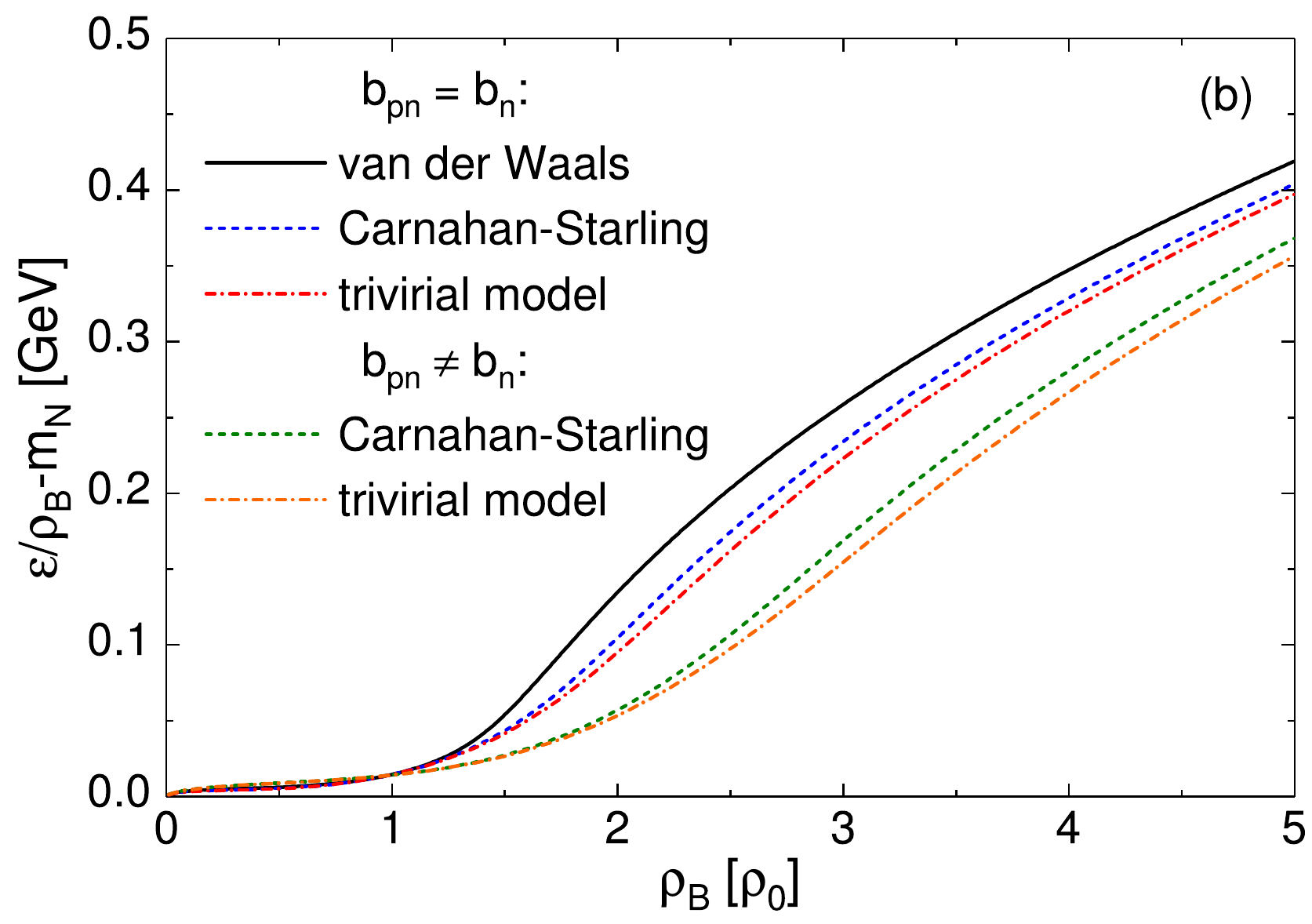}
    \includegraphics[width=.49\textwidth]{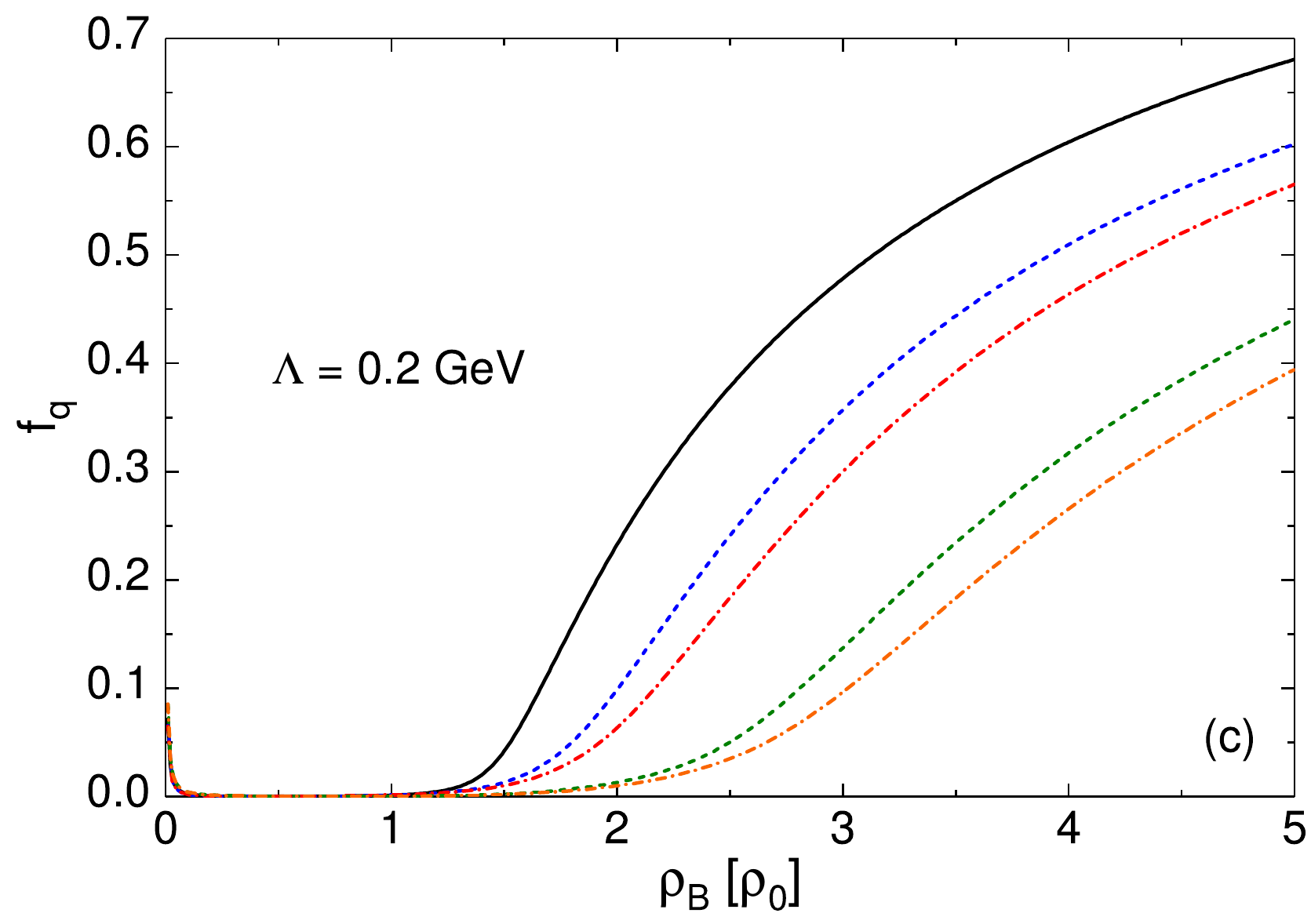}
    \includegraphics[width=.49\textwidth]{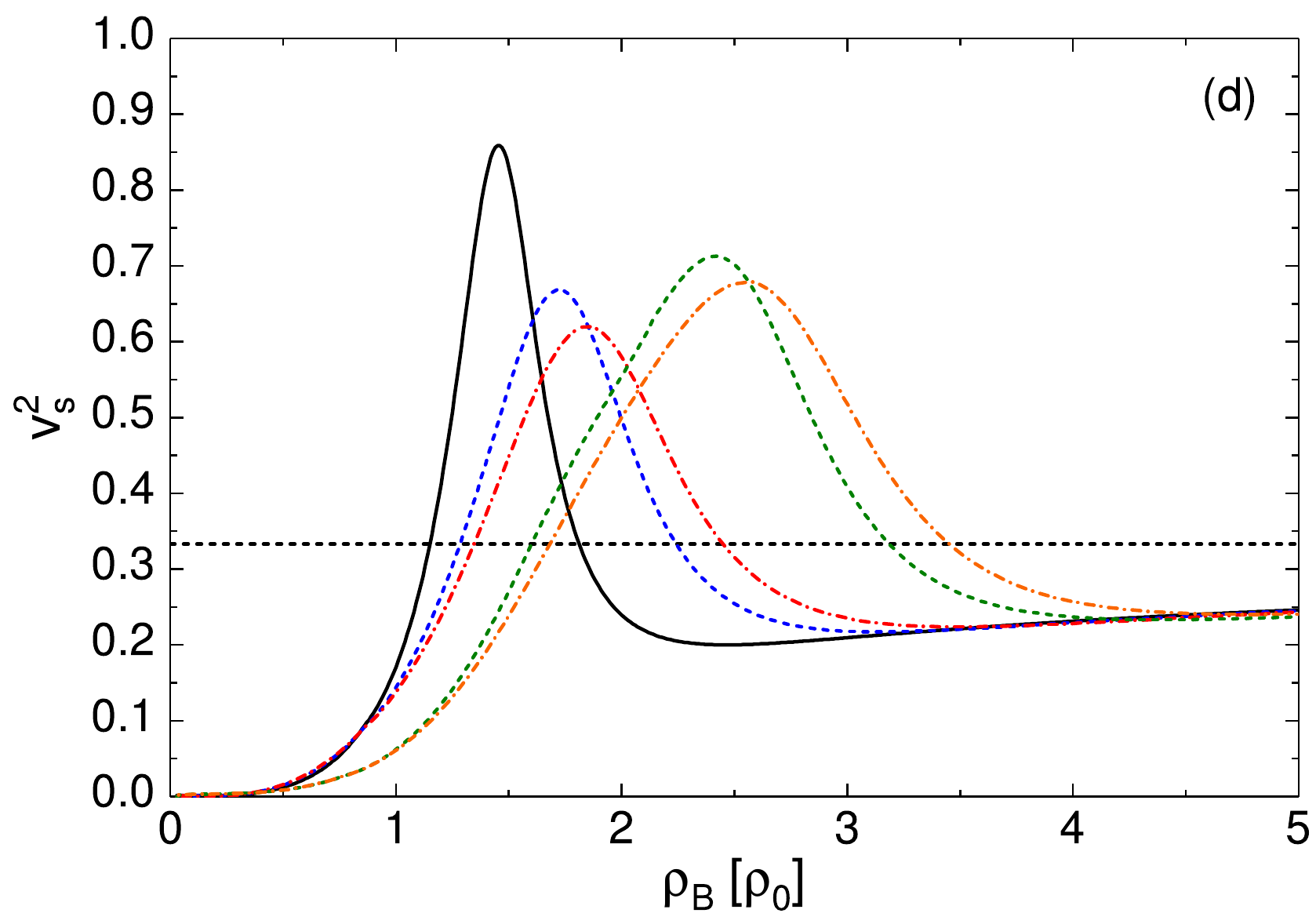}
    \caption{Proton fraction (a), quark fraction (b), energy per baryon (c), and sound velocity (d) as functions of baryon density for quarkyonic star matter with the van der Waals (solid black lines), Carnahan-Starling (dashed blue lines) and trivirial model (dashed-dotted red lines) equation of state. The case $b_{pn}\neq b_n$ is presented for Carnahan-Starling (dashed green lines) and trivirial model (dashed-dotted orange lines) equation of state. Triangles in pallet (a) mark the boundaries of the baryon density interval in which muons are present, with open and filled triangles representing the appearance and disappearance of muons, respectively.}
    \label{fig:td}
\end{figure*}

\section{Neutron star matter}
\label{sec:NS}

\subsection{Beta equilibrium}

To calculate the neutron-star matter equation of state we include lepton contributions and impose beta equilibrium conditions.
The densities of electrons and muons
read
\eq{
\rho_{e,\mu}=\frac{2 (k_F^{e,\mu})^3}{6 \pi^2}~,
}
where $k_F^e$ and $k_F^\mu$  are the Fermi momenta of electrons and muons, respectively.
The density of electrons and muons balances that of protons and quarks to satisfy the charge neutrality condition
\eq{
\rho_e(k_F^e) + \rho_\mu(k_F^\mu)= \rho_Q(y,\rho_B).
}
where $\rho_Q(y,\rho_B) = y \rho_B$. Additionally, at equilibrium $\mu_e=\mu_\mu$. 
The leptonic states are treated here as a noninteracting gas of fermions, therefore, $\rho_{e,\mu}(k_F^{e,\mu}) = (k_F^{e,\mu})^3 / (3\pi^2)$.
For the same reason, the lepton chemical potentials are straightforwardly related to their Fermi energies, $\mu_{e,\mu} = \sqrt{m_{e,\mu}^2 + (k_F^{e,\mu})^2}$,
implying that 
the electron and muon Fermi momenta are related by 
\eq{\label{kfmu}
k_F^{\mu}=\Theta\left[(k_F^e)^2+m_e^2-m_\mu^2\right]\sqrt{(k_F^e)^2+m_e^2-m_\mu^2}
}
where $\Theta$ is the Heaviside step function.
The electron and muon mass used here are $m_e=0.511$~MeV and $m_{\mu}=105.66$~MeV. 

To determine the values of $y$ and $k_F^e$ at given baryon density $\rho_B$ we impose the beta equilibrium condition on the chemical potentials of leptons and electric charge,
\eq{
\mu_Q = -\mu_e.
}
Here
\eq{
\mu_Q = \left(\frac{\partial \varepsilon_{\rm qy} }{\partial \rho_Q} \right)_{\rho_B},
}
where $\varepsilon_{\rm qy}$ stands for the energy density carried by nucleons and quarks (but not leptons).
Taking into account that $\rho_Q = y \rho_B$, one obtains through the chain rule the following
\eq{
\mu_Q = \frac{1}{\rho_B} \, \left(\frac{\partial \varepsilon_{\rm qy} }{\partial y} \right)_{\rho_B}.
}

Therefore, the proton fraction $y$ and electron Fermi momentum $k_F^e$ are determined at given baryon density $\rho_B$ from a system of equations
\eq{
\frac{1}{\rho_B} \, \left(\frac{\partial \varepsilon_{\rm qy} }{\partial y} \right)_{\rho_B} & = -\sqrt{m_e^2 + (k_F^e)^2}, \\\label{ch-neutrality-2}
y \rho_B & = \frac{1}{3\pi^2} \left[(k_F^e)^3 + (k_F^\mu)^3 \right]
}
where $k_F^\mu$ is given by Eq.~(\ref{kfmu}).
This system of equations is solved numerically at each baryon density to determine the charge fraction and lepton Fermi surfaces, and then the quark fraction is subsequently found by energy minimization.

\subsection{Properties of Neutron Stars}

\begin{figure*}[t]
    \includegraphics[width=.49\textwidth]{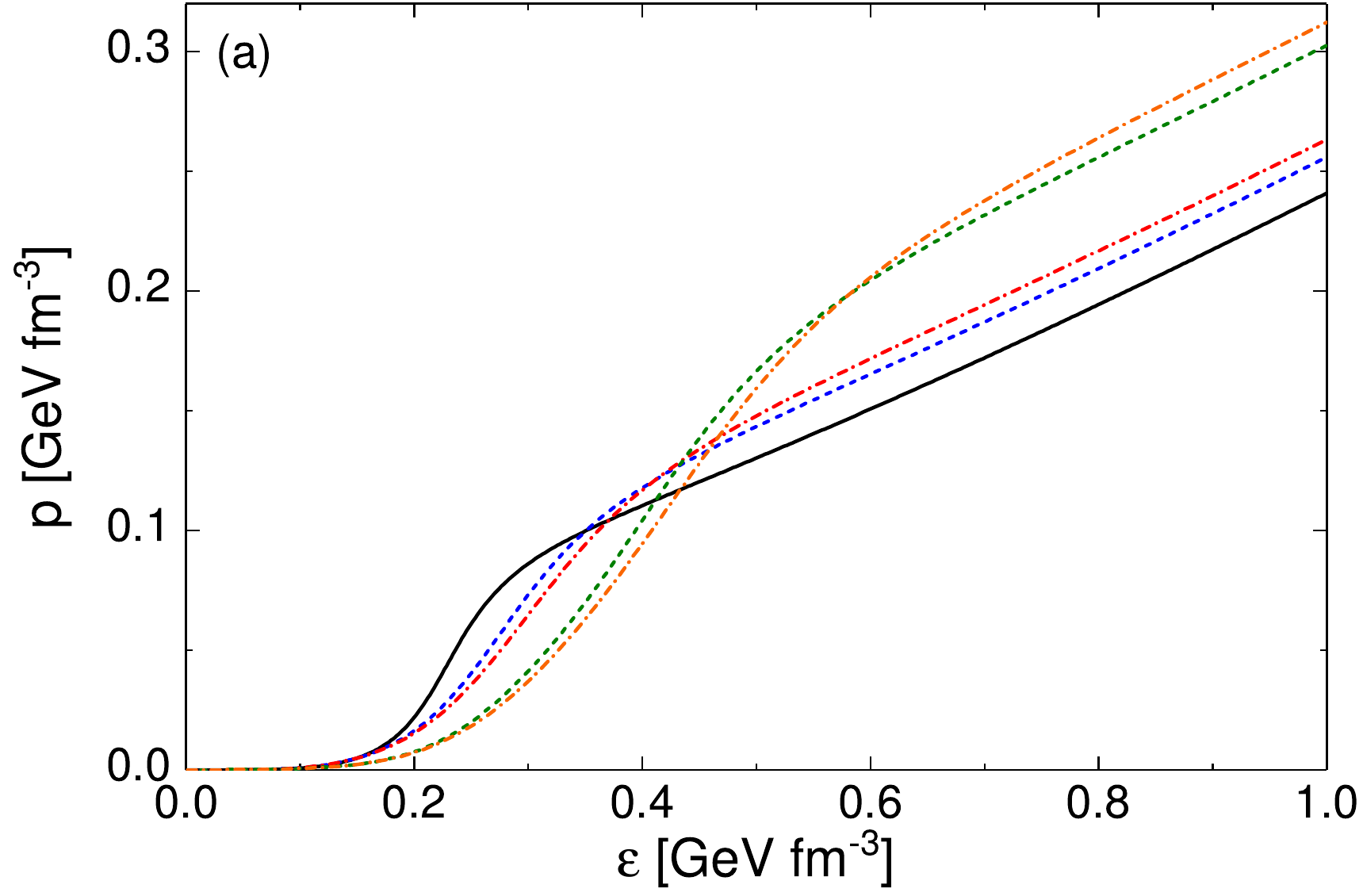}
    \includegraphics[width=.49\textwidth]{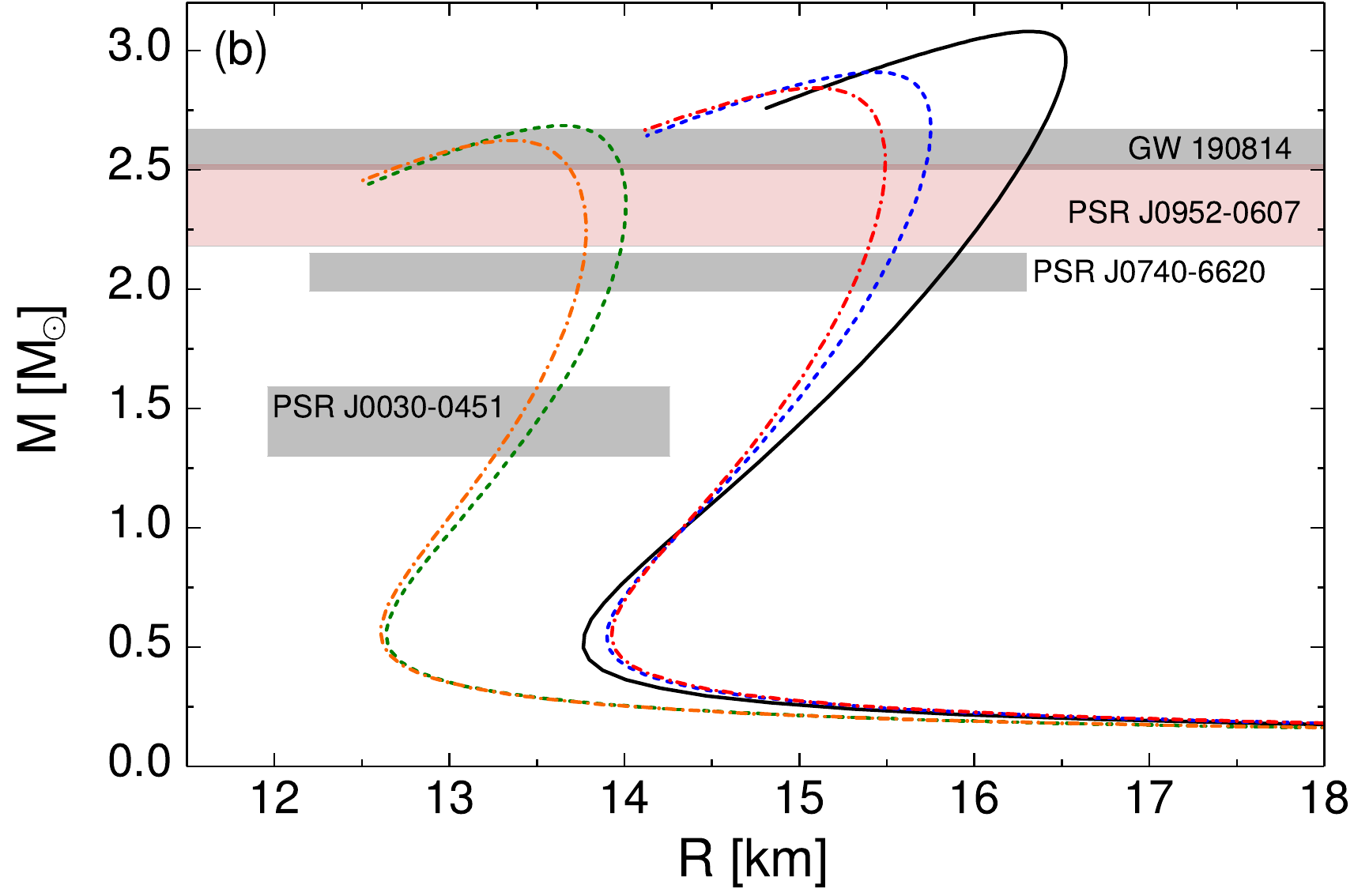}
    \caption{Pressure as a function of energy density (a) and mass-radius relations (b) for quarkyonic star matter with the van der Waals, Carnahan-Starling and trivirial model equation of state. The shaded bands represent the mass constrains for the observations of PSR J0030-0451~\cite{Miller:2019cac,Riley:2019yda} PSR J0740-6620~\cite{Miller:2021qha,Salmi:2022cgy}, PSR J0952-0607~\cite{Romani:2022jhd}, and GW~190814~\cite{Laskos-Patkos:2023tlr}.} 
    \label{fig:NS}
\end{figure*}

The charge fraction $y$ (calculated using the described equilibrium constraints) is shown in Fig.~\ref{fig:td}(a) as a function of baryon density. The interval of $\rho_B$ in which muons are present is marked by triangles on the corresponding curves.
At low densities, we see a rising charge fraction for all excluded volume types. This leads to a rising electron density, causing the chemical potential of electrons to breach the threshold for muon production around $\rho_B\approx 0.8 \rho_0$.

In all cases, the charge fraction exhibits a peak at intermediate densities, $\rho_B\approx 1.4$ -- $2.2 \rho_0$.
Beyond this peak, the behavior differs significantly between cases of $b_{pn}=b_{n}$ and \neqb. 
In \neqb~case, a peak in the charge fraction occurs around $y=0.07$ at $\rho_B\approx 1.4 \rho_0$ before declining to near zero for both Carnahan-Starling and trivirial model excluded volumes.
This drop causes a brief disappearance of muons at $\rho_B\approx 2$ -- $2.5 \rho_0$, which then reappear at higher densities of $\rho_B\approx 4.5$ -- $5 \rho_0$. 
In contrast, when $b_{n}=b_{pn}$ the charge fraction reaches a considerably higher peak, occuring at $y \approx 0.17$ -- $0.19$ and $\rho_B\approx 2$ -- $2.2 \rho_0$ for Carnahan-Starling and  trivirial model excluded volumes, and $y \approx 0.13$ and $\rho_B\approx 1.7 \rho_0$ for van der Waals model. 
Additionally, in this case, the charge fraction remains significantly elevated at large baryon densities with muons consistently present following their initial appearance.

The key features of the equation of state are shown in Fig.~\ref{fig:td}(b) by the 
excess energy per baryon, $\varepsilon/\rho_B - m_N$, with the nucleon mass $m_N = 938$ MeV. The interaction parameters in all considered cases were fixed to reproduce $\left. \varepsilon/\rho_B - m_N \right|_{\rho_B = \rho_0} = -16$~MeV for symmetric nuclear matter which corresponds to the binding energy in the ground state. However, for strongly asymmetric matter considered here the excess energy per baryon is positive at all $\rho_B$ and, correspondingly, the ground state is absent.

Figures~\ref{fig:td} (c) and (d) show, respectively, the quark fraction $f_q\equiv n_q/\rho_B$ where $n_q$ is the contribution of quarks to baryon density and sound velocity $v_s^2$ as functions of $\rho_B$. 
Because the transition density to quarkyonic matter $\rho_{tr}$ is regulated primarily by repulsion $b$, for the case \eqb~the behavior of $f_q$ as a function of $\rho_B$ is similar to the case of symmetric matter (see Ref.~\cite{Poberezhnyuk:2023rct}) in all considered interaction scenarios. 
In this case, the quark fraction increases earlier and more rapidly than in the case of \neqb.
This results in a more rapid stiffening of the equation of state, and leads to a narrower peak in the speed of sound located at density of $\rho_B\approx 1.5-1.8 \rho_0$. 
In contrast, for the case of \neqb~the strength of neutron-neutron repulsion is significantly lower, $b_{n}\approx 0.86 b$ for CS and $b_{n}\approx 0.81 b$ for TVM. 
As a result, the onset of quarks takes place at higher baryon densities and is more gradual.
In all cases, one observes a non-zero quark fraction at very low baryon densities, $\rho_B \ll \rho_0$.
This has been discussed in Ref.~\cite{Sen:2020qcd} as an artifact due to the infrared regulator $\Lambda$, which can be mitigated by introducing density dependence into $\Lambda$. Since the artifact is only relevant at low densities, it has negligible overall impact on the equation of state.

In order to compare our model predictions with observational data we must calculate the neutron star equation of state, i.e. pressure as a function of energy density. This EoS is then used as input to the Tolman-Oppenheimer-Volkoff (TOV) equation to compute neutron star properties such as mass-to-radius ratios and tidal deformability. 

The pressure is computed through the Euler relation,
\eq{
\varepsilon + p = Ts + \mu_B \rho_B + \mu_Q \rho_Q + \mu_e n_e + \mu_\mu n_\mu.
}
The last three terms cancel each other out due to charge neutrality and beta-equilibrium conditons, and the entropic contribution vanishes when evaluated at $T=0$.
Therefore,
\eq{\nonumber
p & = -\varepsilon + \mu_B \rho_B \\\label{pressure}
& = -\varepsilon_{\rm qy} - \varepsilon_e - \varepsilon_{\mu} + \mu_B \rho_B
}
where baryochemical potential $\mu_B$ is obtained as $\mu_B=(\partial\varepsilon/\partial\rho_B)_{\rho_Q}$.
One can also write the pressure as
\eq{
p = p_{\rm qy} + p_{e} + p_{\mu},
}
where
\eq{
p_{\rm qy} & = -\varepsilon_{\rm qy} + \mu_B \rho_B + \mu_Q \rho_Q, \\
p_{e,\mu} & = -\varepsilon_{e,\mu} + \mu_{e,\mu} \rho_{e,\mu}.
}

Figures~\ref{fig:NS} (a) and (b) present, respectively, pressure as a function of energy density (\ref{pressure}) and quarkyonic star mass-radius curves for both considered scenarios. 
To obtain mass-radius relations we use the TOV equation solver~\cite{Motornenko:2019arp,TOVsolver}, which merges the input EoS (\ref{pressure}) with the EoS for the neutron star crust at low baryon densities and calculates the resulting mass-radius relations. 
We see that in \neqb~scenario quarkyonic matter can support neutron stars with masses up to $M \simeq 2.6 M_{\odot}$ and radii up to $R\simeq 14$~km.
Because of the earlier stiffening, the \eqb~ case supports a higher maximal radius and mass of quarkyonic star.

We compare the obtained mass-radius curves with constraints from observations of PSR J0030-0451~\cite{Miller:2019cac,Riley:2019yda} PSR J0740-6620~\cite{Miller:2021qha,Salmi:2022cgy}, PSR J0952-0607~\cite{Romani:2022jhd}, and GW~190814~\cite{Laskos-Patkos:2023tlr}.
One sees that the model aligns well with these constraints for the \neqb~parameter set that reproduces the lower estimates of the symmetry energy slope.
It also implicitly favors the compact object of mass $M \simeq 2.6 M_{\odot}$ observed in the GW~190814 event~\cite{LIGOScientific:2020zkf} to be a neutron star~(see Refs.~\cite{Most:2020bba,Tan:2020ics,Fattoyev:2020cws} for the discussion of various possibilities).
The case \eqb, on the other hand, predicts too large radii~($R > 14$~km) of $1.4 M_{\odot}$ neutron stars disfavored by PSR J0030-0451, and arguably too high maximum masses of $M \simeq 2.8-3.0 M_{\odot}$. 

We also calculate normalized tidal deformabilities $\Lambda$ for the case of trivirial model excluded volume, again with TOV equation solver~\cite{Motornenko:2019arp,TOVsolver}. We evaluate at $M=1.4 M_{\odot}$ for comparison with observational constraints from GW170817 and GW190814. In the case of \neqb~ we find $\Lambda \simeq 800$, consistent with the upper limits of tidal deformability for GW170817~\cite{LIGOScientific:2017vwq} and other analysis of GW190814~\cite{LIGOScientific:2020zkf}. In contrast, the case of \eqb~ results in much higher values of $\Lambda \simeq 1500$. For this scenario of isospin-blind excluded volume we find comparable values of $\Lambda \simeq 800$ at higher neutron star masses of $M=1.6 M_{\odot}$.

\section{Summary}
\label{sec:summary}

In this work we have investigated Quarkyonic Matter properties for arbitrary charge fraction in the range $0 < y < 1/2$ at zero temperature. Nuclear interactions were incorporated via the van der Waals formalism, and interaction parameters were constrained using symmetry energy, slope of symmetry energy, and ground-state nuclear properties. We then expanded this framework to a general real gas model by utilizing the excluded-volume formalisms of Carnahan-Starling and the Trivirial Model. This led to a more stable behavior in the speed of sound, namely peaks which remained causal for all values of isospin asymmetry. The speed of sound was seen to rise above the conformal limit as density increased during the hadronic phase. The onset of deconfined quarks then softens the equation of state, bringing the speed of sound back below the conformal limit before approaching this limit from below at high densities. We also investigated the cases of isospin-blind \eqb~and isospin-dependent \neqb~repulsion parameter, finding notably different behavior for each case regardless of the excluded-volume formalism used.

In order to apply this framework to the neutron star equation of state, we included leptons under the constraints of charge neutrality and beta decay equilibrium. This fixes the system's charge fraction at a given baryon density and causes a rising charge fraction at low baryon densities during the purely hadronic phase, which then peaks and becomes suppressed within the quarkyonic phase for higher densities. 
We then utilized the Euler relation to compute the pressure of the system as a function of energy density, giving us a complete equation of state of neutron-star matter.

This EoS was used as input to the TOV equation to compute mass-radius relations for neutron stars. 
We find that the case of \eqb~leads to larger stars at a given mass compared to the \neqb~case, and larger maximum masses.
The case \neqb~gives a closer match to observational constraints.

We note many areas for further progress, which have been left for future work. 
We aim to extend the model to finite temperatures, likely requiring an explicit treatment of quark-hadron duality between the quark and hadron distribution functions. 
This will allow us to employ the EoS to neutron star mergers as well as heavy-ion collisions.
We also note that including strange degrees of freedom could be important for describing neutron star matter and would require a parameterization of nucleon-hyperon interactions and appropriate treatment of the Pauli exclusion principle for the strange quark~\cite{Fujimoto:2024doc}. 
Finally, we note that our van der Waals parameterization overestimates nuclear incompressibility $K_0$. 
We plan to resolve the issue by including density-dependent attraction terms. 
Lastly, we note our choice of the infrared regulator to be $\Lambda = 200$ MeV, chosen to regulate the divergent speed of sound characteristic of Quarkyonic Matter. In practice, this is a free parameter of the quasi-particle approach to quarkyonic matter, which determines how sharp the peak in $v_s^2$ is, but not at which density it occurs.
This parameter could be included as a density-dependent term or contrained with a Bayesian analysis as was explored recently~\cite{Pang:2023dqj}.

\begin{acknowledgments}

R.P. acknowledges the support from 
the Philipp Schwartz Initiative of the Alexander von Humboldt Foundation.

\end{acknowledgments}

\bibliography{main}

\end{document}